# Transport Mechanism of Acetamide in Deep Eutectic Solvents

H. Srinivasan[1], V.K. Sharma[1], V.García Sakai[2], Jan P Embs[3], R. Mukhopadhyay[1,4] and S. Mitra[1,4*]

[1]Solid State Physics Division, Bhabha Atomic Research Centre, Mumbai 400085, India

[2]ISIS Facility, Science and Technology Facilities Council, Rutherford Appleton Laboratory, Didcot, U.K.

[3]Laboratory for Neutron Scattering, Paul Scherrer Institute, Villigen, Swizerland

[4]Homi Bhabha National Institute, Anushaktinagar, Mumbai 400094, India

* Corresponding author: smitra@barc.gov.in

Tel:+91-22-25594674; FAX:+91-22-25505151

# Abstract

Over the last couple of decades, deep eutectic solvents (DESs) have emerged as novel alternatives to ionic liquids that are extensively used in synthesis of innovative materials, metal processing, catalysis, etc. However, their usage is limited, primarily because of the large viscosity and poor conductivity. Therefore, an understanding of the molecular origin of these properties is essential to improve their industrial applicability. Here, we present the report of the nanoscopic diffusion mechanism of acetamide in a DES synthesized with lithium perchlorate as studied using neutron scattering and molecular dynamics (MD) simulation techniques. Although, the acetamide based DES (ADES) has remarkably lower freezing point compared to pure acetamide, the molecular mobility is found to be enormously restricted in the former. MD simulation indicates a diffusion model with two distinct processes, corresponding to, long range jump diffusion and localised diffusion within a restricted volume. This model is validated by analysis of neutron scattering data in both molten acetamide and ADES. The long range diffusion process of acetamide is slower by a factor of three in ADES in comparison with molten acetamide. MD simulation reveals that the long range diffusion in ADES is restricted mainly due to the formation of hydrogen bond mediated complexes between the ionic species of the salt and acetamide molecules. Hence, the origin of higher viscosity observed in ADES can be attributed to the complexation. The complex formation also explains the inhibition of the crystallisation process while cooling and thereby results in depression of the freezing point of ADES.

# 1. Introduction

In the last couple of decades, there has been a growing interest in the study of ionic liquids (ILs). This has been encouraged by their wide scope in industrial applications, like metal processing[1,2], batteries[3], biomass treatment[4], super-capacitors[5], carbon dioxide capture[6,7] and many more. Despite these advances, the poor biodegradability[8] and hazardous toxicity[8,9] of the ILs pose some serious problems. Further, their synthesis is not very eco-friendly, as huge amounts of salts and solvents form an important requisite in that process. These setbacks along with their high production costs form a major hindrance to widespread commercial and industrial use of ILs. The beginning of this century saw the emergence of a novel variety of solvents, namely deep eutectic solvents (DESs)[10-16], which were found to possess physicochemical properties very similar to ILs. Apart from their cheaper production costs they are also environmentally more benign. DES are formed by the mixture of two or more compounds at a particular molar ratio corresponding to their eutectic point[11,15,16]. Generally, the mixtures are found to have a significantly lower freezing point compared to the parent compounds[15,16]. These solvents have found great applicability in various industrial processes like electrodeposition[13,17-20], catalysis[21], nanoparticle[21] and nanotube[22,23] synthesis, drug transport[24], $CO_2$ capture[25], etc. In most of the aforementioned applications, DESs have been found as a more eco-friendly and cheaper alternative to ILs. The high solubility of metal salts, coupled with better conductivity compared to non aqueous solvents, make DES a very viable alternative in the metal deposition industry[15,17]. It has been observed that certain drugs insoluble or poorly soluble in water showed thousand times better solubility in DES[24]. A new kind of catalysts in the form of gold nanowire networks were synthesised using the choline chloride/urea (reline) DES[21]. Multiwalled carbon nanotubes have also been synthesised using the same DES[23]. It has been shown that reline is a promising candidate for catalyst in various reactions and shown to be better alternative to ILs and lipases[26-28].

Generally, eutectic mixtures are homogenous mixtures of two or more components with a freezing point significantly lower than that of their individual components. This depression in freezing point is directly related to the interaction strength between the components in the mixture, such that, stronger interaction between them leads to a larger depression of the freezing point. A schematic phase diagram of a two component eutectic mixture is shown in Fig. 1. The composition at which the lowest freezing point is achieved is called the eutectic point. The depression in freezing point, $\Delta T_f$, which is the difference between the freezing temperature at the eutectic point and the mean freezing temperature of the mixture, is indicated in the figure. In the case of DES, they are usually formed by mixing quaternary ammonium salts and metal salts, where the former acts as hydrogen bond donor and the latter as an acceptor[15]. The charge delocalisation due to hydrogen bonding in these mixtures is believed to be the key reason for depression in freezing point[11,15]. Further, the low lattice energy of the large non-symmetric ions of quaternary ammonium salts is favourable to attain lower freezing points of the DES[15].

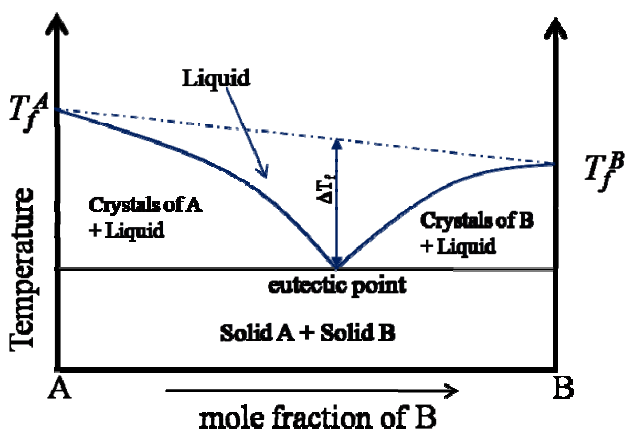

**Figure 1.** *Schematic phase diagram of a two component eutectic mixture.*

Recently, it has been shown that a mixture of acetamide ($CH_3CONH_2$) and a group of lithium salts (LiX, X = $ClO_4$, Br, $NO_3$) in a molar ratio of 78:22 form DES having freezing points below room temperature[29]. It has been suggested that depression of freezing points in these systems can be ascribed to the ability of these salts to break the inter-amide hydrogen bonding. Among the three salts, $LiClO_4$ forms the least viscous DES at room temperature, indicating larger affinity of perchlorate ion to amide group[29].

Acetamide in its liquid phase has the capability to dissolve a plethora of organic and inorganic compounds due to its large dipole moment and dielectric constant[30]. Naturally, liquid acetamide has great applicability as a solvent media in various industrial processes. Therefore, having molten mixtures of acetamide at significantly lower temperatures is of immense importance to industrial processes. DES based on other lithium salts like LiTFSI and LiPF$_6$ with N-methylacetamide have been shown to be candidates of electrolytes in lithium ion batteries.[31] Certain mixtures of acetamide and lithium salts (LiTFSI, LiClO$_4$) have shown promising features as electric double layer capacitors due to their excellent electrochemical stability [32]. Further, similar DESs have been put to use in various industrial processes, like metal processing, electrodeposition, etc. The transport properties of a DES, particularly conductivity and viscosity of the liquid mixture, play a crucial part in these applications. It has been observed that the conductivity of a DES is inversely correlated with its viscosity, which has been found to be significantly larger for a majority of DES at ambient temperature. This remains an important setback for the effective application of DES as an alternative to ILs.

While there are plenty of studies involving the macroscopic transport of DES systems, a comprehensive model of microscopic dynamics is not well established. Although, phenomenological descriptions of molecular mobility have been suggested from their macroscopic transport properties, direct experimental studies on their microscopic dynamics are yet to be probed in detail. Among the phenomenological models, hole-theory, which had been used to describe the dynamics of molecules in ionic mixtures and molecular liquids, was found to explain the behaviour of viscosity in a variety of DESs.[10,33] The model is well established for DESs with lower viscosity where the anion is assumed to move together with the amides; but in the case of DESs with higher viscosity it works only if the motion of amide is assumed to be independent of the anion.[10] However, nuclear magnetic resonance (NMR) experiments on choline chloride based DES have shown that diffusion of the molecules is not controlled by the formation of holes, but a free volume model is more appropriate.[34] Yet again, the free volume model was proposed as phenomenological description to the results observed in NMR spectroscopy.

Quasielastic neutron scattering (QENS) spectroscopy exploits the wavelength and energy of thermal neutrons to probe the diffusion mechanism of the system at lengths of few Angstroms and detect different relaxation processes in timescale of picosec to nanosec[35-37]. It is particularly suitable to obtain information about the geometrical and dynamical features of diffusion processes in hydrogenated systems. Some recent QENS studies on pyridinium based ionic liquids have revealed two independent relaxation processes in the system[38,39]. Similarly, different mobilities of different molecules in glyceline DES (glycerol+choline chloride) was also observed using QENS experiments[40]. Therefore, it is an excellent probe to observe the dynamics of acetamide in DES at a molecular scale. On the other hand, classical molecular dynamics (MD) simulation technique provides atomistic insights about the system which are generally inaccessible to experimental methods[36,41]. MD simulation has been used to study various properties in DES, including their carbon dioxide solubility and effects of hydrogen bond strength[42,43]. Here, we report the nanoscopic dynamics of molten-acetamide (hereafter referred as ACM) and an acetamide based DES made with lithium perchlorate in the molar ratio of 78:22 (hereafter referred as ADES) as studied using QENS and MD simulation techniques.

## 2. Materials and Methods

Acetamide ($CH_3CONH_2$, 99% purity) and lithium perchlorate ($LiClO_4$, 98% purity) were obtained from Sigma Aldrich. A solid mixture of acetamide and lithium perchlorate in the molar ratio of 78:22 was heated at a temperature of 340 K for approximately 2 hours, until a clear solution of the ADES was formed. The mixture remained in the liquid phase after cooling it down to room temperature (300 K).

QENS experiments were carried out using the IRIS spectrometer at the ISIS neutron and muon facility at RAL, UK. IRIS is a backscattering spectrometer which uses a pyrolitic graphite. Using the instrument in the offset mode with (002) reflection, the energy resolution of the instrument is ~17.5 μeV (Full width at half maximum) and the energy transfer range is -0.3 to 1.0 meV. The wave-vector ($Q$) transfer range of the spectrometer is between 0.5 – 1.8 Å$^{-1}$. QENS experiments on acetamide were carried out at 300 and 365 K, and for DES mixtures they were carried out over a range of temperatures – 300, 315, 330, 355 and 365 K. The lowest temperature was chosen considering the eutectic temperature of ADES (~290 K) and the highest temperature was chosen to accordingly compare with molten acetamide. Since a large quasielastic broadening was observed in the case of molten acetamide at 365K, especially at higher Q-values, QENS experiments for the same samples were also carried out at the FOCUS spectrometer (with wavelength=6Å) at the Paul Scherrer Institute (PSI), Switzerland, providing wider energy transfer range. In the used configuration, FOCUS spectrometer has an energy resolution of ~ 45 µeV. Standard vanadium sample was used to measure the energy resolution of both spectrometers.

MD simulation of ACM and ADES was also carried out. The initial configuration of both the systems were constructed by randomly arranging molecules and ions in a cubic box using Packmol[44]. In the case of the former, 512 acetamide molecules were used in the simulation, and for ADES, 400 acetamide molecules and 56 lithium ions and perchlorate ions were used resulting in the molar ratio 78:22. The CHARMM force field (version 27)[45] was used for acetamide and the parameter set for perchlorate and lithium ions were taken from the work of Lopes & Deschamp[46] and Joung & Cheatham[47] respectively. The system was equilibrated for 10 ns in the NPT ensemble, using Langevin barostat and thermostat with a target pressure of 1 atm and a target temperature of 365 K. Subsequent to temperature and system density equilibration, the simulation was continued in the NPT ensemble for another 5 ns, to record atomistic trajectories at an interval of 1 ps for further analysis. A separate simulation was also carried out to record short trajectories with a smaller time interval of 0.01 ps for 100 ps. The MD simulations were carried out using DL-Poly-4[48].

## 3. Theoretical aspects

In general for a neutron scattering experiment, the intensity of the scattered neutrons is given by the double differential scattering cross section which is expressed as a sum of two individual contributions arising from coherent and incoherent scattering of neutrons from the sample and can be written as[35],

$$\frac{d^2\sigma}{dEd\Omega} \propto \frac{k_f}{k_i}[\sigma_{coh}S_{coh}(\mathbf{Q},E) + \sigma_{inc}S_{inc}(\mathbf{Q},E)] \tag{1}$$

where, $\sigma_{coh}$ and $\sigma_{inc}$ are the coherent and incoherent scattering cross sections, $S_{coh}$ and $S_{inc}$ are the coherent and incoherent scattering laws; $E$ is the energy transfer and $\mathbf{Q} = \mathbf{k}_f - \mathbf{k}_i$, is the momentum transfer in the scattering process. It should be noted that neutron scattering spectra from a hydrogenous sample is dominated by incoherent scattering from the hydrogen atoms in the sample, due to its exceptionally high incoherent scattering cross section ($\sigma^H_{inc} >> \sigma^{an\ atomy}_{inc/coh}$). In that case, eq. (1) can be written as

$$\frac{d^2\sigma}{dEd\Omega} \propto \frac{k_f}{k_i}\sigma_{inc}S_{inc}(\boldsymbol{Q},E) \qquad (2)$$

The incoherent scattering law is related to the van-Hove self-correlation function, $G_s(\boldsymbol{r},t)$, through a space-time Fourier transform[35],

$$S_{inc}(\boldsymbol{Q},E) = \int_{-\infty}^{\infty} dt \int d^3r \ e^{-i(Et-\boldsymbol{Q}\cdot\boldsymbol{r})} G_s(\boldsymbol{r},t) \qquad (3)$$

The correlation function, $G_s(\boldsymbol{r},t)$, gives the probability of finding a particle at a position $\boldsymbol{r}$ after a time $t$, if it started at the origin at $t = 0$. The incoherent scattering law observed for the case of liquids, $S_{inc}(Q,E)$, is averaged all the $Q$-orientations due to the isotropic nature of the system. It provides both geometrical and temporal information regarding the stochastic motion of individual atoms (predominantly of hydrogen atoms) in the system. However, since the experimental spectra are an ensemble average over the system, the information about the system can only be extracted from phenomenological modelling and the individual components of motion in the system cannot be discerned, in general. For this purpose, we resort to the technique of classical MD simulation, where the atomistic details of the trajectories enable us to isolate the individual components of the motion. MD simulations can be used as a guide to choose an appropriate model to analyse the QENS spectra of the system. Further, it is computationally viable to carry out MD simulations, for time and length scales of motions directly accessible in QENS experiments, making the combination of the two techniques a very powerful to gain molecular information of the system.

**4 Results and discussion**

First we develop a model for acetamide diffusion based on the results of MD simulation on ACM and ADES at 365 K. The model is then verified by analysing the QENS data of both the systems at 365 K. Subsequently, the same model is employed for the QENS data of ADES at lower temperatures to study the thermal evolution of its dynamical parameters. Finally, the molecular origin of the diffusion mechanism in these systems is discussed

*4.1 Development of a dynamical model – MD simulation*

MD simulation of ACM and ADES were carried out at 365 K using an all-atom model and information about each individual atomic trajectory was extracted. The trajectories were analysed with the aim of developing a physical model to describe the diffusion of acetamide molecules in the system, which can be subsequently used to aid the analysis and interpretation of the QENS experimental spectra. The dynamics of molecules in a liquid can be separated into two fundamental degrees of freedom, the motion of molecular centre of mass (COM) and the motion of the atoms with respect to the COM. We shall refer to the latter as internal motion of the molecule. At sufficiently high temperatures, where the interaction energies are expected to be weaker compared to the thermal energies, an assumption of decoupling of these two motions is a fairly good approximation. Given the temperatures chosen in this work, the individual components of motion were analysed with independent models.

*4.1.1 Dynamics of acetamide COM*

To begin with, the COM trajectories of the acetamide molecule were calculated over the entire range of simulation time (5 ns). The mean-squared displacement (MSD) of the COM of acetamide explicitly calculated from its trajectory is shown in Fig. 2 for ACM and ADES. At very short times ($t < 0.1$ ps), a quadratic dependence corresponding to the ballistic regime is observed for both the systems. In this regime, the particle motion remains unhindered by collisions or interactions and hence remains the same for both the systems. This is followed by an almost plateau like region showing a transient sub-diffusive behaviour. This is generally observed in hydrogen-bonded liquid systems which show transient caging and subsequent diffusion after H-bond breaking. The extent of this plateau region is slightly longer in the case of ADES compared to ACM, indicating that H-bond interaction is stronger in this system. Eventually in both the cases, the asymptotic region of MSD shows simple linear dependence with time, indicating a Brownian motion. A linear fit (Fig. 2) of MSD in this asymptotic limit, using the Einstein's relation, MSD = $6D_{com}t$, yields the free diffusion coefficient of the acetamide molecules. In the case of ACM, the $D_{com}$ it is found to be, $0.91 \times 10^{-5}$ cm$^2$/s, which is at least 3 times larger than that found for the case of ADES (~ $0.29 \times 10^{-5}$ cm$^2$/s). While the MSD is a useful starting point and reveals the asymptotic behaviour of the diffusion of acetamide molecules, a more comprehensive picture can be obtained by calculating intermediate incoherent scattering function (IISF). The IISF corresponding to COM can be calculated using the following formula,

$$I_{COM}(Q,t) = \frac{1}{N}\sum_{j=1}^{N} \overline{\left\langle e^{-i\mathbf{Q}.\mathbf{R}_j(t_0)} e^{i\mathbf{Q}.\mathbf{R}_j(t+t_0)} \right\rangle} \quad (4)$$

where, $\mathbf{R}_j(t)$ is the position of COM of the $j^{th}$ molecule, $t_0$ is an arbitrary time-origin and $N$ is the total number of acetamide molecules in the system. The average over time-origins, $t_0$, is denoted by angular-brackets and the bar indicates the average over all $Q$-orientations to represent isotropic conditions in the liquid system. It is to be noted that the periodic boundary condition used in the simulation restricts the value of $Q$ as given by $Q = (2\pi n/L)$, where $L$ is the length of the cubic box used in the MD simulation and $n$ is an integer.

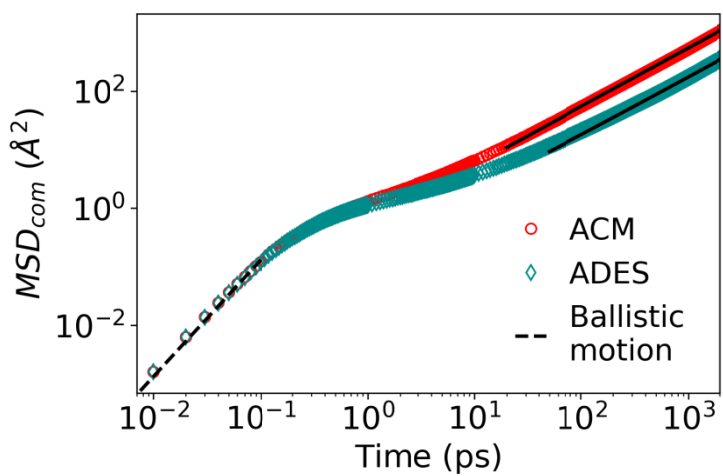

**Figure 2.** *MSD of acetamide COM in ACM and ADES calculated from the MD simulation trajectory. The linear fits (free diffusion) in the asymptotic limit and quadratic fit (ballistic motion) at small times (t < 0.1 ps) are also shown in the plot.*

The physical model for the IISF of the COM is chosen considering three different processes – ballistic motion, localized motion inside a transient cage and cage-cage free diffusion. This is motivated by observing the trajectories of the molecular COM shown in Fig. 3 for ACM (left) and ADES (right), in which local clustering and jump-like motion are visible. A similar model has been employed in studying the dynamics of supercooled water by Qvist et al.[49]. Since the ballistic motion of the molecules is temporally well separated from its diffusive processes, the IISF for acetamide COM can be written in the following fashion,

$$I_{com}(Q,t) = \alpha(Q) I_{diff}(Q,t) + (1-\alpha(Q)) e^{-(\mu_v Qt)^2/2} \qquad (5)$$

where, $\alpha$ is the weight factor associated with diffusive processes, $I_{diff}(Q,t)$ is the IISF for the two diffusive processes (localised diffusion, cage-to-cage free diffusion) and $\mu_v$ is the average thermal velocity of the acetamide COM which is equal to $(k_B T/m)^{1/2}$, where $m$ is the molecular mass of acetamide. The diffusion processes involved in this model, although statistically independent, need not be fully temporally separated[49] and hence $I_{diff}(Q,t)$ takes following form,

$$\begin{aligned} I_{diff}(Q,t) &= I_{free}(Q,t) I_{loc}(Q,t) \\ &= e^{-\zeta_1 t}\left[C_0(Q) + (1-C_0(Q)) e^{-\zeta_2 t}\right] \end{aligned} \qquad (6)$$

where the first term corresponds to free diffusion process (cage-cage jump diffusion) and the second term in the square-brackets corresponds to the localised motion within the transient cage with $\zeta_1$ and $\zeta_2$ as their respective relaxation rates. $C_0(Q)$ is called the structure factor corresponding to localized motion – it's also referred to as the elastic incoherent structure factor (EISF) as it represents the elastic contribution in a QENS experiment arising due to a localized motion. Therefore, IISF corresponding to COM motion (eq. 5) can be written as,

$$I_{com}(Q,t) = \alpha(Q)\left[C_0(Q) e^{-\zeta_1 t} + (1-C_0(Q)) e^{-(\zeta_1+\zeta_2)t}\right] + (1-\alpha(Q)) e^{-(\sigma_v Qt)^2/2} \qquad (7)$$

where the first term in the square brackets correspond to the diffusive components of motion which include diffusion inside the transient cage and cage-cage free diffusion. The last term described by a Gaussian term represents the ballistic motion of the molecule. The least-squares fits of eq. (7) to the calculated $I_{com}(Q,t)$ are shown in Fig. 4 (a) for ACM and ADES at $Q = 1.2$ Å$^{-1}$. The individual components are also indicated in these figures. The diffusive components are indicated as component 1 & 2 and component 3 represents the ballistic motion. The quality of fits is found to be excellent which confirms the validity of the model. Fig. 4(b) shows the comparative plot of $I_{com}(Q,t)$ for both the systems, clearly indicating that the COM motion is significantly slower in ADES compared to ACM. The solid lines in Fig 4(b) represent their respective model fitting functions based on eq. (7).

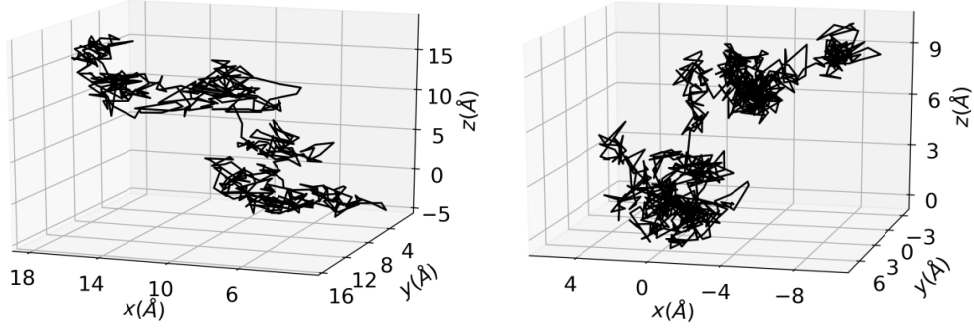

**Figure 3.** *COM trajectory of a given acetamide molecule in (left) ACM and (right) ADES from MD simulation.*

Firstly, the variation of $\zeta_1$ over the entire $Q$-range, is analysed in order to validate the proposition of the cage-to-cage jump diffusion process associated with the acetamide COM. Fig. 5(a) shows the plot of $\zeta_1$ with respect to $Q^2$, as calculated from the fits of eq. (7) for the IISF of the COM for the two systems. It is clear from Fig. 5 (a) that the jump diffusion is markedly slower in ADES compared to ACM. The jump diffusion of the acetamide COM is modeled using a Singwi-Sjolander (SS) model[50], which assumes the jump lengths to be completely random. In the SS model, the equation describing the relaxation rate $\zeta_1$ is given by,

$$\zeta_1(Q) = \frac{D_j Q^2}{1 + D_j Q^2 \tau_j} \tag{8}$$

where $D_j$ is the diffusion coefficient associated to cage-cage jump diffusion process and $\tau_j$ is the mean residence time of the particle between subsequent jumps. The asymptotic nature of the jump diffusion described by this equation can be seen in the limit $Q \to 0$, where $\zeta_1 \to D_j Q^2$ corresponding to Fickian diffusion law. The fitting of $\zeta_1$ using eq. (8) is shown in Fig. 5 (a) and the parameters obtained from the fit for acetamide COM at 365K, are listed in Table 1. The diffusivity of acetamide in ADES is smaller by a factor of ~ 3 in comparison to ACM. Further, the longer mean residence time for the COM in ADES is an indication of stronger H-bonding network of the acetamide with the ionic species of the salt. In both cases, the jump diffusion coefficient matches well with the diffusivity obtained in the asymptotic limit from the MSD.

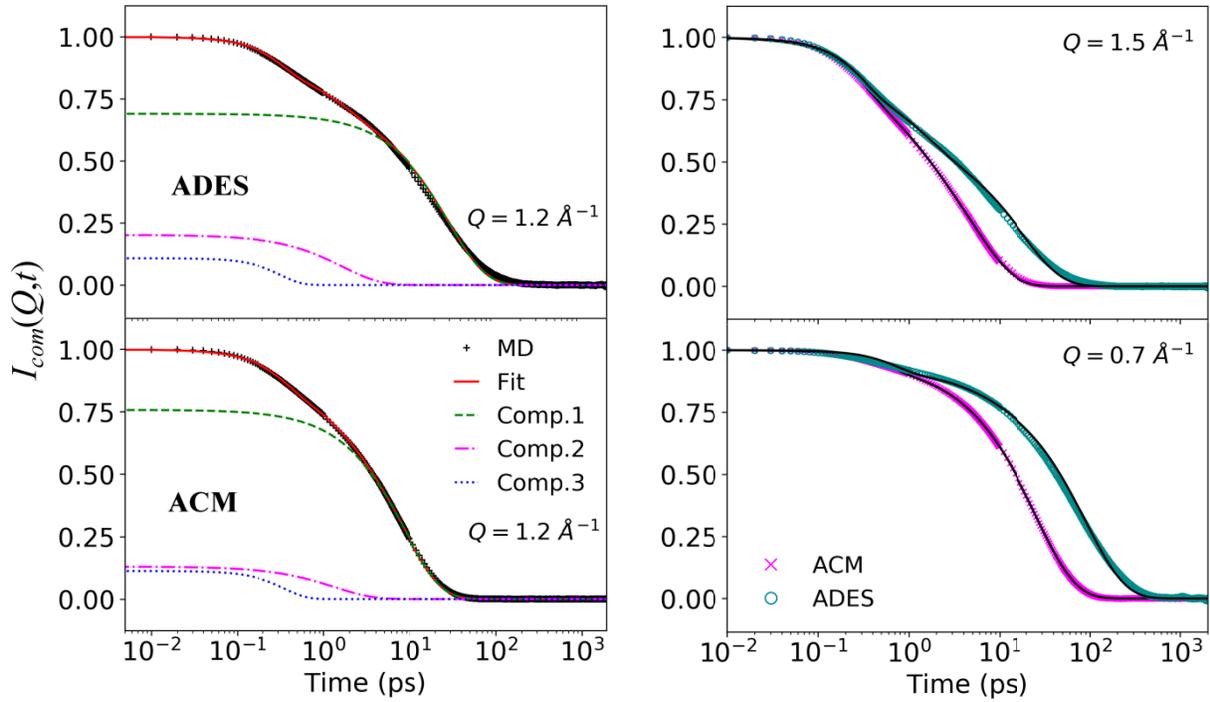

**Figure 4.** *(a) $I_{com}(Q,t)$, calculated using eq. (4) from the MD simulation trajectory of acetamide COM for ADES and ACM. The fits based on eq. (7) along with their components are shown in the figure. The comp. 1 and comp. 2 correspond to the diffusive components and comp. 3 corresponds to the ballistic motion. (b) Comparative plots $I_{com}(Q,t)$, for ACM and ADES at two typical Q-values are shown. The solid black lines represent the respective fits based on eq. (7).*

The parameters $C_0$ and $\zeta_2$ (from the fits of eq. (7)) corresponding to the structure factor and relaxation rate of localised motion obtained from least-square fits of the IISF are, shown in Fig. 5(b). It is evident that the local dynamics of the COM is not significantly different in both the cases. In order to model the localised translation inside the transient cage, the localised translational diffusion (LTD) model within a sphere of radius $r$ was considered. In this model the IISF of a particle diffusing within a cage of radius $r$ with a diffusivity $D_{loc}^{COM}$ is given by[51],

$$I_{LTD}(Q,t) = A_0^0(Qr) + \sum_{l \neq 0, n \neq 0}(2l+1)A_n^l(Qr)e^{-\frac{(x_n^l)^2 D_{COM}^{loc}}{r^2}t} \tag{9}$$

where the first term arises due to the localized nature of the motion and contains information about its geometry. In QENS spectra, this information is contained in the elastic peak, called EISF, as defined earlier in eq. (6). The second term is the sum over all the contributions to the quasielastic part, which contains information about the dynamical nature of the motion. The model IISF, $I_{LTD}(Q,t)$ can be compared to $I_{loc}(Q,t)$ given in eq. (6) to validate the nature of the localised motion. However prior to employing this model, based on inspection of the trajectories of individual COM (Fig. 3), it is found that the radii of the clusters are distributed, rather than having a single value. In this case, we consider the case of an exponential random distribution for the radii of the spheres and therefore modify the model function for the IISF of LTD to,

$$I_{LTD}^{mod}(Q,t) = \int_0^\infty dr \; p(r) I_{LTD}(Q,t) \tag{9a}$$

where $p(r)$ is an exponential distribution function with average radius of $r_{avg}$ and is given by,

$$p(r) = \frac{1}{r_{avg}} e^{-r/r_{avg}} \tag{9b}$$

The corresponding elastic incoherent structure factor (EISF), $A_0^{0(mod)}(Q)$, which describes the geometry of the motion for the modified LTD model then becomes[51],

$$A_0^{0(mod)}(Q) = 3\int_0^\infty dr \left[\frac{j_1(Qr)}{Qr}\right]^2 p(r) \tag{9c}$$

This equation is used to fit, $C_0(Q)$ with $r_{avg}$ as a fitting parameter. The fitting of the model function given by eq. (9c) is shown in the inset of Fig. 5(b); the average radius, $r_{avg}$ is found to be 0.8 and 0.9 Å for ACM and ADES respectively. Using the value of $r_{avg}$, the second term arising from the sum of exponential relaxation functions, can be computed numerically for different values $D_{loc}^{COM}$ at each $Q$. The corresponding model relaxation rate for each $Q$-value is evaluated from the inverse of time $t$ at which the total sum is equal to $e^{-1}$. The variation of $\zeta_2$ with respect to $Q$ and its calculated curve based on the modified LTD model are shown in Fig. 5 (b). The localized diffusion coefficient for the COM of acetamide from MD simulation at 365 K for ACM and ADES is listed in Table 1. It is observed from the obtained set of parameters that the local dynamics of the COM doesn't show appreciable difference between the ACM and DES.

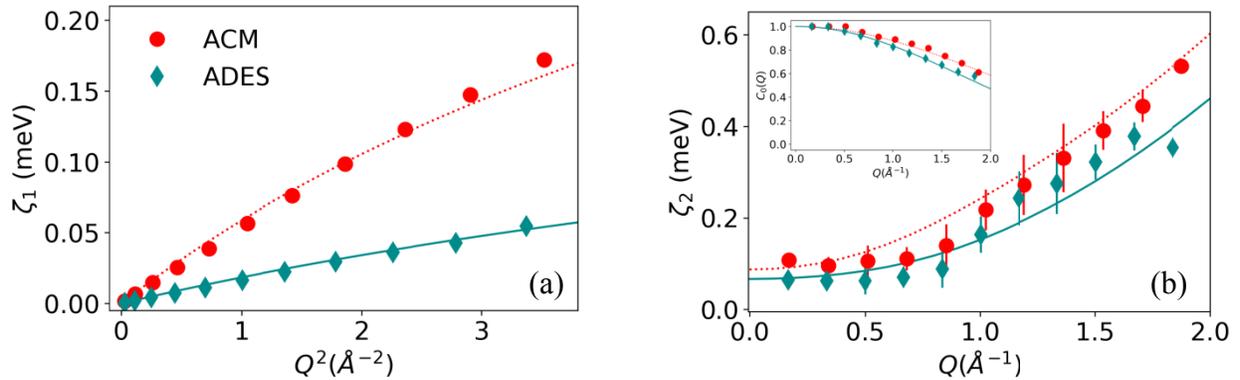

**Figure 5.** *(a) Variation of $\zeta_1$ with respect $Q^2$. The dotted and solid lines are according to Singwi-Sjolander (SS) jump diffusion model for ACM and ADES respectively. (b) Variation of $\zeta_2(Q)$ for ACM and ADES; the dotted and solid lines show the calculated model function (as described in the text). The inset shows the variation of the EISF of the localized diffusion, $C_0(Q)$ along with lines showing the fitted model functions for the modified LTD (eq. 9c).*

*4.1.2 Internal dynamics of acetamide*

The motion of acetamide atoms observed in the QENS experiments is a superposition of the dynamics of the molecular COM as well as the average of dynamics of the individual hydrogen atoms with respect to the COM. Having investigated the motion of COM exclusively, we now turn to motion of H-atoms with respect to the COM which is referred to as the internal motion. The trajectories of H-atoms in the COM frame of each acetamide molecule are calculated over the entire simulation time (5 ns) and is used to calculate the IISF of internal motion using the following formula,

$$I_{int}(Q,t) = \frac{1}{N_H}\sum_{j=1}^{N_H} \overline{\left\langle e^{-i\mathbf{Q}\cdot\mathbf{d}_j(t_0)} e^{i\mathbf{Q}\cdot\mathbf{d}_j(t+t_0)} \right\rangle} \tag{10}$$

where $\mathbf{d}_j(t)$ is the position of $j^{th}$ hydrogen atom with respect to it's molecular COM and $N_H$ is the total number of hydrogen atoms in a molecule. The angular brackets and the bar denote the same meaning as before in eq. (4). The motion of H-atoms with respect to the COM in liquid systems is generally well described by isotropic rotation on a sphere. However, the trajectories of the hydrogen atoms about the COM encompass a spherical shell, after sufficiently long time (~ 200 ps). At shorter times, fluctuations about the mean position on the surface of the sphere most probably arise from fast conformational dynamics. Therefore, the IISF for internal motion was fit using the following model function[35],

$$\begin{aligned}I_{int}(Q,t) &= I_{iso}(Q,t)I_{fast}(Q,t) \\ &= \left( j_0(Qa)^2 + \sum_{l=1}^{\infty}(2l+1)j_l(Qa)^2 e^{-l(l+1)D_R t} \right)\left( b_0(Q) + (1-b_0(Q))e^{-\chi t} \right)\end{aligned} \tag{11}$$

where the first term corresponds to isotropic rotational motion[35] and second to the fast conformational motions. As both motions are localised, each contains a structure factor associated to their dynamics. While it is represented by $b_0(Q)$ for the fast motions, it is given by $j_0(Qa)^2$ for the isotropic rotation. In eq. 11, $a$ is the radius of the sphere encompassed by the isotropic rotation, $D_R$ is the rotational diffusion constant and $\chi$ is the relaxation rate of the fast conformational dynamics. The fitting of $I_{int}(Q,t)$ was done considering upto $l=10$, since higher $l$ terms give a negligible contribution in the relevant $Q$-range. The calculated $I_{int}(Q,t)$ and its respective fits using eq. (11) are shown in Fig. 6 for both ACM and ADES, where the individual components from isotropic rotation and fast motions are also indicated. The $Q$-averaged value of the radius of rotation, $a$, for ACM and ADES is found to be 2.0 and 2.2 Å, which is in good agreement with the mean distance of atoms from the acetamide COM. However, the rotational diffusivity, $D_R$, is 0.04 and 0.02 ps$^{-1}$ for ACM and ADES respectively, indicative of a substantially slower rotational dynamics of acetamide in ADES compared to ACM at 365 K. The observed values of $D_R$, are comparable to the rotational diffusivity of ethylene glycol in bulk at 348 K (~0.03 ps$^{-1}$), which is also a similarly H-bonded liquid[52]. The relaxation process corresponding to conformational dynamics for both the systems was found to be very fast, with the variation of χ between 2-6 meV. Therefore, this motion is not expected to contribute to the QENS spectra observed in both the spectrometers.

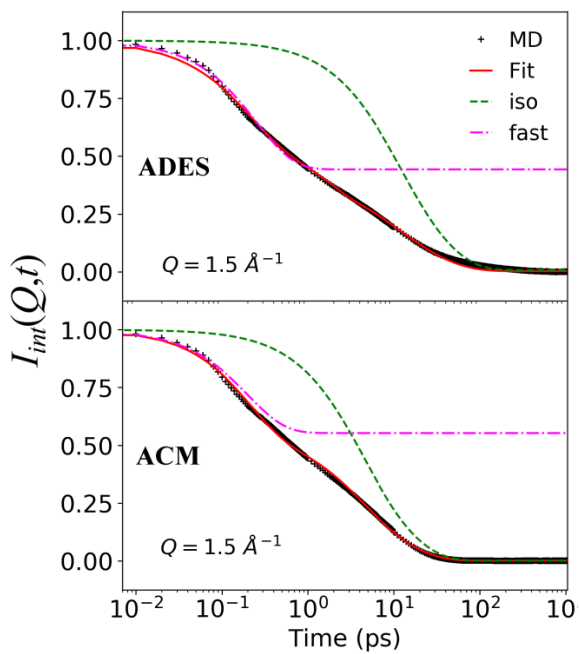

**Figure 6.** $I_{int}(Q,t)$, calculated using eq. (10) from the MD simulation trajectory for ACM and ADES at a given Q-value. The fits based on eq. (11) along with their components are also shown.

**Table 1.** *List of dynamical parameters obtained from the model described in the text for ACM and ADES from QENS and MD simulation at 365 K.*

| System | | $D_j(10^{-5}\ cm^2/s)$ | $\tau$ (ps) | $r_{avg}$ (Å) | $D_{loc}^{COM}(10^{-5}\ cm^2/s)$ | $D_{loc}^{H}(10^{-5}\ cm^2/s)$ |
|---|---|---|---|---|---|---|
| ACM | MD | 1.00 (±0.06) | 1.3 (±0.1) | 0.8 (±0.1) | 2.81 (±0.08) | - |
| | QENS | 0.92 (±0.02) | 1.8 (±0.4) | 1.8 (±0.1) | - | 2.16 (±0.08) |
| ADES | MD | 0.30 (±0.02) | 2.9 (±0.1) | 0.9 (±0.1) | 2.51 (±0.06) | |
| | QENS | 0.29 (±0.02) | 3.3 (±0.2) | 1.7(±0.2) | - | 1.36 (±0.05) |

### 4.2 Validation of the dynamical model – QENS data analysis

The QENS spectra of acetamide and ADES (acetamide + lithium perchlorate) measured at the IRIS spectrometer are shown in Fig. 7 at a typical $Q$-value of 1.2 Å$^{-1}$ for two different temperatures (300 K and 365 K). At 300 K, significant quasielastic broadening is observed in the case of ADES (Fig. 7(a)) indicating the presence of stochastic motion of acetamide in the system; whereas no observable broadening is found to be present in the case of pure acetamide suggesting that no molecular motions are detected within the resolution of the spectrometer. This is likely because, at 300 K, pure acetamide remains in the solid phase where the diffusion of acetamide molecules is not observable in the accessible time scale of the spectrometer. However above the melting point of acetamide ($M_p$) ~ 352 K, (at 365 K) quasielastic broadening for molten-acetamide is observed to be significantly larger compared to that of ADES (Fig. 7(b)). The width of the quasielastic broadening is a direct measure of the mobility in the system. Therefore, it is clear that acetamide molecules are significantly more mobile in their pure molten state than in the ADES (at 365 K). This is in qualitative agreement with the results of MD simulation of ACM and ADES at 365 K. It is interesting to note that, although ADES has a much lower freezing point compared to pure acetamide, the molecular dynamics of molten-acetamide (ACM) is much faster compared to ADES at any given temperature above $M_p$. A study of individual components of molecular diffusion in the both systems above $M_p$ will reveal the effects of the salt (lithium perchlorate) on the local and global dynamics of acetamide.

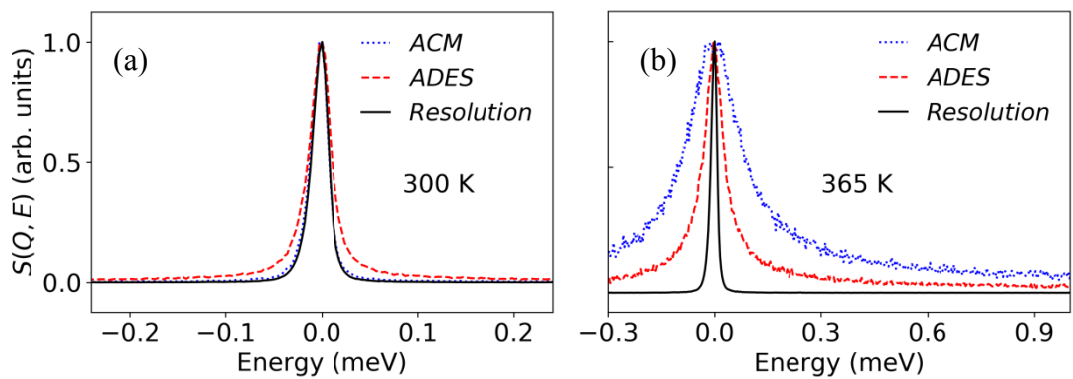

**Figure 7**. *QENS spectra of acetamide and ADES, for a Q-value of 1.2 Å$^{-1}$ at (a) 300 K and (b) 365 K measured at IRIS spectrometer. The instrumental resolution measured using a vanadium stanadard is shown by the solid lines. All spectra are peak normalised to the unity for direct comparison of QE broadening.*

The dynamical model developed based on MD simulations of these systems shall be used to analyse the QENS data of ACM and ADES at 365 K. In the case of ACM, all data from the FOCUS spectrometer is considered, only the lower $Q$-values are analysed from IRIS as the quasielastic broadening at high $Q$ was too large for the accessible energy transfer window. The model used for the analysis was found to be consistent with the experimental data from both spectrometers. However, all the QENS data of ADES used in the analysis are measured at IRIS spectrometer.

As pointed out earlier, the measured QENS signal is dominated by scattering from H-atoms, and hence we observe their motion alone. The dynamics of H-atoms can arise from the COM and the internal motion of the molecules. The simulation results indicated that, there are two relaxation processes involved in the motion of COM – jump diffusion and localized diffusion within transient cages; and it also showed that their relaxation-rates lie in the energy transfer range of 0.05 to 1 meV, which is within the observable range of the QENS spectrometers used in this work. It should also be noted that the two timescales are well separated and can be resolved well with the QENS spectrometers used. MD simulations also showed that the internal motions of acetamide encompass two relaxation processes – isotropic rotational diffusion and fast conformational dynamics. The relaxation rate of the fast motion lies in the energy transfer range of 2-6 meV, which lies outside the range of both IRIS and FOCUS. In the case of the isotropic rotation, an effective relaxation rate can be estimated by approximating the sum of exponentials as a single exponential function. The effective relaxation rate has an energy transfer range between 0.2 to 0.5 meV. This causes a significant overlap between the range of relaxation timescales associated with the localized motion of the COM and isotropic rotation of H-atoms. Therefore, it is not possible to differentiate these two dynamical processes from the analysis of the QENS spectra. In order to circumvent this problem, we propose an alternatively simpler dynamical model for the motion of H-atoms, which takes into account all the features observed from MD simulation. In this model, we consider two dynamical processes, (a) jump diffusion of acetamide COM and (b) localised translation diffusion of hydrogen atoms. The first process can be directly compared with the jump diffusion of acetamide COM which was observed in the analysis of the simulation trajectories. Process (b) can be reckoned as an effective motion which arises as a superposition of localized diffusion of COM and isotropic rotation of the H-atoms about the COM. The scattering law pertaining to the model described above can be written mathematically as the convolution of scattering laws associated with the two processes,

$$S_{model}(Q,E) = L_j(\Gamma_j, E) \otimes S_{loc}(Q,E)$$
$$= L_j(\Gamma_j, E) \otimes [A_0 \delta(E) + (1-A_0) L_{loc}(\Gamma_{loc}, E)] \quad (12)$$

where the $L_j$ and $L_{loc}$ are Lorentzians corresponding to jump diffusion and localised diffusion processes with the half-width half-maxima (HWHM), $\Gamma_j$ and $\Gamma_{loc}$ respectively. $A_0$ is the elastic incoherent structure factor (EISF) associated to the localized diffusion process, which possesses information about the spatial profile of the local motion. To compare with the QENS data one has to convolute the model function with the resolution of the spectrometer, The scattering law used to describe the quasielastic data then can be written as,

$$S(Q,E) = \left[ A_0(Q) L_j(\Gamma_j, E) + (1-A_0(Q)) L_{j+loc}(\Gamma_j + \Gamma_{loc}, E) \right] \otimes R(E) \quad (12a)$$

where, $R(E)$ is the resolution of the spectrometer. The measured QENS spectra of molten acetamide along with the fits based on eq. (12a) for ACM and ADES are shown in Fig. 8 at a typical $Q$ of 1.4 Å$^{-1}$. The individual components corresponding to Lorentzians, $L_j$ and $L_{j+loc}$, are also shown in the figure.

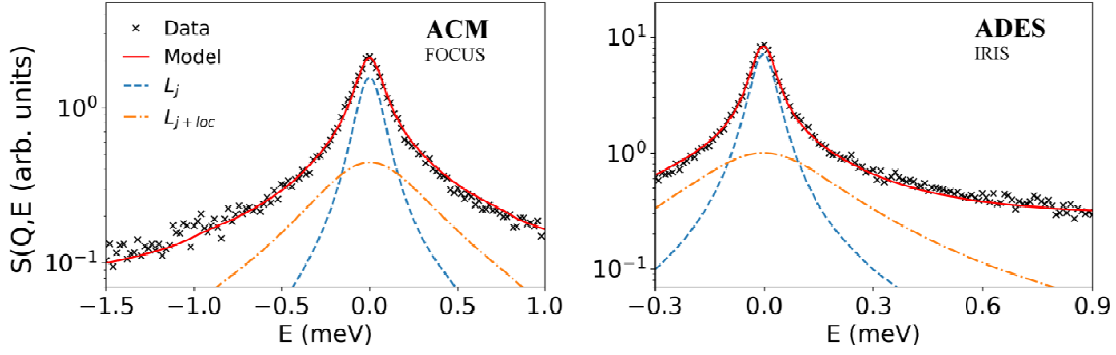

**Figure 8.** *QENS spectra of ACM and ADES measured at FOCUS and IRIS spectrometer respectively, shown at a typical Q-value of 1.4 Å$^{-1}$. The individual components of the model (eq. 12a) are also indicated.*

The Singwi-Sjolander[50](SS) model of jump diffusion is used in the investigation of the process (a), which is the jump diffusion of acetamide COM. This is the same model that has been already employed to explain the jump diffusion of COM from MD simulation trajectories. The model function for the variation of $\Gamma_j$ is similar to the one given in eq. (8),

$$\Gamma_j(Q) = \frac{D_j^{ex} Q^2}{1 + D_j^{ex} Q^2 \tau_j^{ex}} \quad (13)$$

where, $D_j^{ex}$ is the diffusion coefficient in the jump diffusion process and $\tau_j^{ex}$ is the mean residence time between subsequent jumps. The parameter, $\Gamma_j$ obtained from fitting the QENS spectra along with the SS model function is shown together for ACM and ADES in Fig. 9 (a). The jump diffusion coefficient $D_j^{ex}$, obtained from the fitting for both the systems are listed in Table 1. The jump diffusivity of acetamide in ADES is found to be smaller by a factor of 3 compared to ACM. This observation is found to be consistent with the results observed in MD simulations. The excellent agreement of parameters (Table 1) between the results of MD simulation and QENS strongly supports the validity of the model. The jump diffusivities found in this study are comparable to that of bulk ethylene glycol at 348 K which is reported with a diffusivity of ~0.6 × 10$^{-5}$ cm$^2$/s and mean residence time of ~ 7 ps[52].

The localised diffusion within a sphere, i.e., process (b), has already been discussed in the context of COM motion in MD simulation (section 4.1.1). Now this is considered in the analysis of QENS spectra which concerns motion of the H-atoms alone. The scattering law associated for diffusion within a sphere of radius $r$ and diffusivity $D_H^{loc}$ is obtained by a time Fourier transform of eq. (9) and is given by,[51]

$$S_{loc}(Q,E) = A_0^0(Qr)\delta(E) + \sum_{l\neq 0, n\neq 0}(2l+1)A_n^l(Qr)\frac{\frac{(x_n^l)^2 D_H^{loc}}{r^2}}{\left(\frac{(x_n^l)^2 D_H^{loc}}{r^2}\right)^2 + E^2} \qquad (14)$$

where first term on the RHS is the elastic component and second is the quasielastic part. Similar to the earlier case of COM motion, an exponential distribution of radii is considered, rather than a single radius and therefore the model function for EISF would be the same as given in eq. (9c). The plot of $A_0$ vs $Q$, obtained from fitting the QENS spectra of liquid acetamide at 365 K using eq. (12a), is shown Fig. 9(b). The fitting of the EISF model function from eq. (9c), for localised diffusion is also indicated by dotted and solid lines for ACM and ADES respectively. The EISF fits quite well with the model considered, yielding an average radius, $r_{avg} \sim 1.8$ Å in both the cases. It may be noted that the average radius of the sphere is about two times larger than that obtained from MD simulation; this is not surprising since it should be viewed as an effective combination of the isotropic rotation of the H-atoms about COM and the localised diffusion of the COM. The variation of $\Gamma_{loc}$ and the associated theoretical model function with respect to $Q$ is shown in Fig. 8 (b). The theoretical model function is obtained from calculating the HWHM of the sum of Lorentzians given in eq. (14). The obtained values of $D_{loc}^{H}$ and $r_{avg}$ are given in Table 1; it is evident from the parameters that, although the geometry of the localised dynamics is not altered, the localised diffusion is slightly slower in ADES compared ACM. This might arise due to the decreased rotational diffusivity in ADES as observed from MD simulations. The fits of $A_0$ and $\Gamma_{loc}$ shown in Fig. 9 (b) suggest that the model describing H-atoms diffusing within spheres describes the QENS data successfully. Further, it is also consistent with the picture of dynamical model inferred from the results of MD simulation (Table 1).

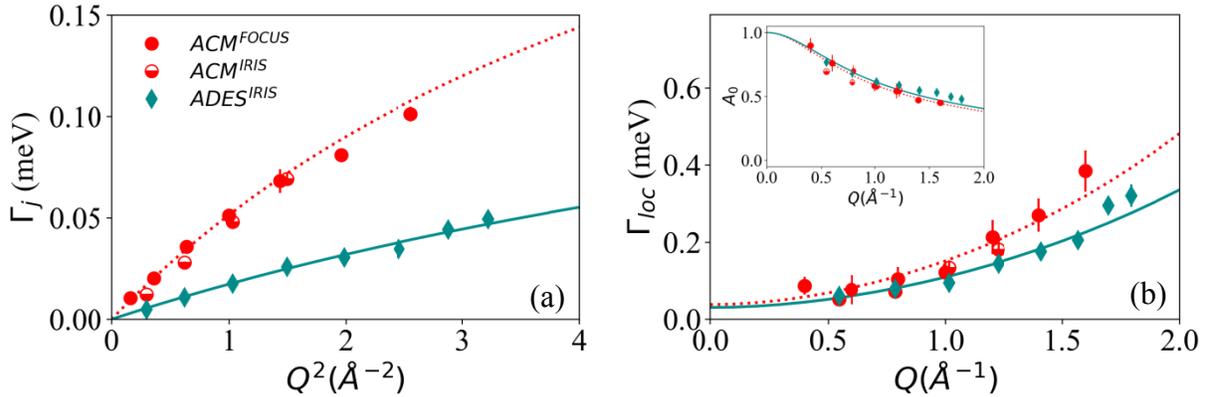

**Figure 9.** *(a) Variation of HWHM of the jump diffusion process, $\Gamma_j$, with respect to $Q^2$ for ACM and ADES at 365 K. The dotted (ACM) and solid (ADES) lines denote the SS model of jump diffusion given by eq. (13). (b) HWHM associated with the localised diffusion, $\Gamma_{loc}$, the calculated model functions are represented by the respective lines. Data from both IRIS and FOCUS spectrometers is shown. The inset shows the EISF, $A_0$, of localised diffusion for molten acetamide, the fits with the model function based on eq. (9c) are also shown.*

*4.3 Thermotropic evolution of acetamide dynamics in ADES– QENS analysis*

The dynamical model as described in the previous section is used to describe the QENS spectra of ADES for all the measured temperatures. The spectra, fits and the individual fit components (eq. 12a) are shown in Fig. 10 for 300, 315 and 330 K. It is evident from the quality of the fits that the dynamical model developed from the results of MD simulation is suitable to describe the dynamics successfully over the entire temperature-range from 300 – 365 K. A plot of the HWHM corresponding to jump diffusion of acetamide, $\Gamma_j$ with $Q^2$ is shown for all temperatures in Fig. 11; the model functions, based on SS model (eq. 13), at each temperature are also indicated by the solid lines in the figure. The values of $D_j^{ex}$ and $\tau_j^{ex}$ for all the temperatures are summarised in Table 2. At 300 K, the diffusivity of acetamide in ADES is comparable to that of cholinium ions in ethaline[34] measured using NMR technique; while in reline and glyceline the diffusion of the cholinium ions are significantly slower [34,40]. The temperature dependence of diffusion constant, $D_j$ follows an Arrhenius temperature dependence given by,

$$D_j = D_{j0} e^{-E_A/k_B T} \qquad (15)$$

where $E_A$ is the activation energy of the diffusion process. The Arrhenius fit of $D_j$ is shown in the inset of Fig. 11. The activation energy for acetamide in ADES considered here is found to be 20.2 (±1.5) kJ/mol, which is comparable to other DES systems given in the literature[34]. The decrease in diffusivity and increase in residence times at lower temperatures can be ascribed to lower thermal energy for the acetamide to move beyond the potential barrier of inter-amide and amide-ion hydrogen bonds.

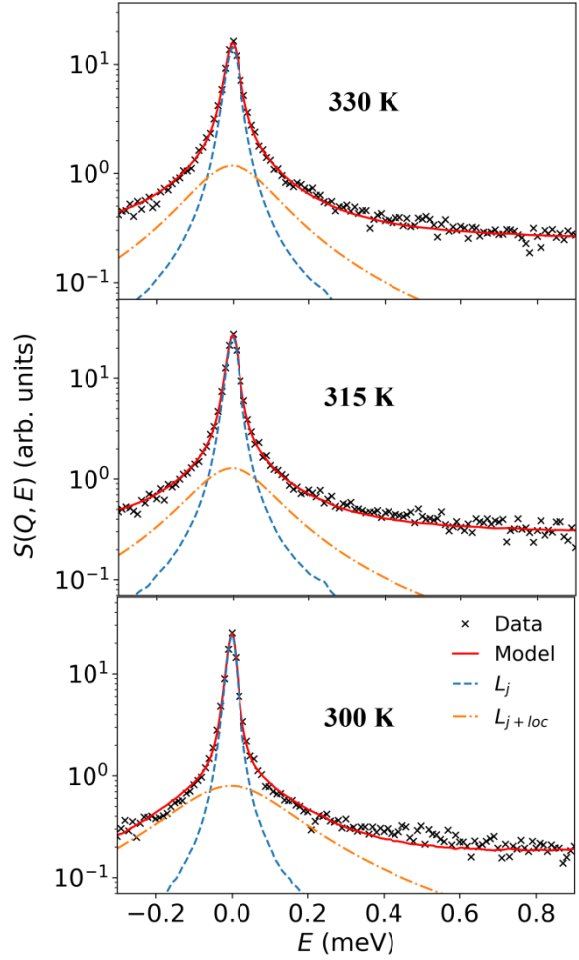

**Figure 10.** *QENS spectra of ADES measured at 300, 315 and 330 K with IRIS spectrometer, shown at a typical Q-value of 1.4 Å$^{-1}$. The individual components of the model (eq. 12a) are also indicated.*

The information about localised motions of acetamide in ADES is contained in two parameters – the average radius ($r_{avg}$) of the confinement and the localised diffusion constant ($D_H^{loc}$). The EISF obtained from the fitting of QENS spectra at different temperatures, is shown in Fig. 12(a). The solid lines in the figure are the respective fits based on theoretical model for localised diffusion within a sphere (eq. 9c). The average radii obtained from these fits at each temperature are given in Table 2; as expected, they clearly show an increasing trend with rising temperature. The obtained values of radii are similar to the recent results on glyecline DES$^{40}$. Similarly, the HWHM associated to the Lorentzian describing the localised diffusion, $\Gamma_{loc}$, within a sphere is shown in Fig. 12(b). The increase in temperature causes increase in the HWHM indicating greater hydrogen mobility. The solid lines in Fig. 12(b) show the calculated function based on the LTD model, at different temperatures. The change in localised diffusion constant, $D_H^{loc}$, from the lowest to highest temperature is found to be only about three times, whereas jump diffusivity changes by an order of magnitude. However, the variation of $D_H^{loc}$ with temperature also follows Arrhenius law, with an activation energy of 14.3 (±0.8) kJ/mol, indicating that localised diffusion is slightly less sensitive to temperature in comparison to jump diffusion. Nevertheless, it is clear that both the spatial and temporal features of the acetamide diffusion in ADES are significantly affected by temperature and in particular, the jump diffusion process is strongly hindered with decreasing temperature.

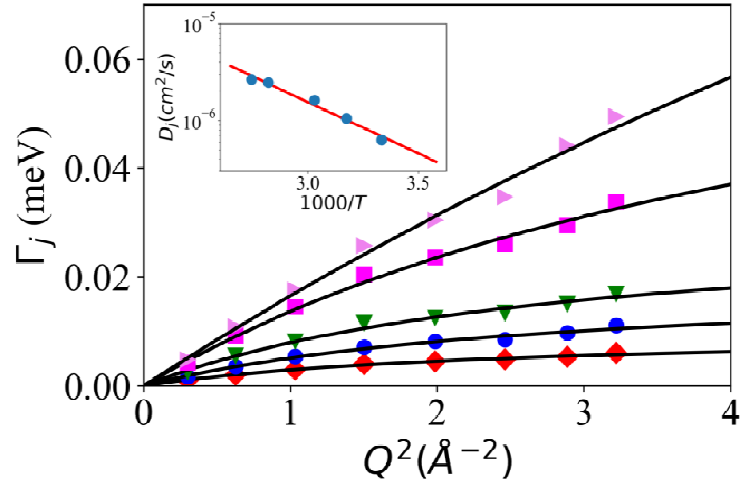

**Figure 11**. *HWHM associated to the Lorentzian describing the jump diffusion of acetamide COM in ADES at all observed temperatures. 365 K(▶), 355 K(■), 330 K(▼), 320 K (●), 300 K (♦). The solid lines indicate the respective fits using eq. (13) based on the SS model. Arrhenius plot of jump diffusion constant, $D_j$ is shown in the inset. The solid line (red) is the fit based on Arrhenius equation.*

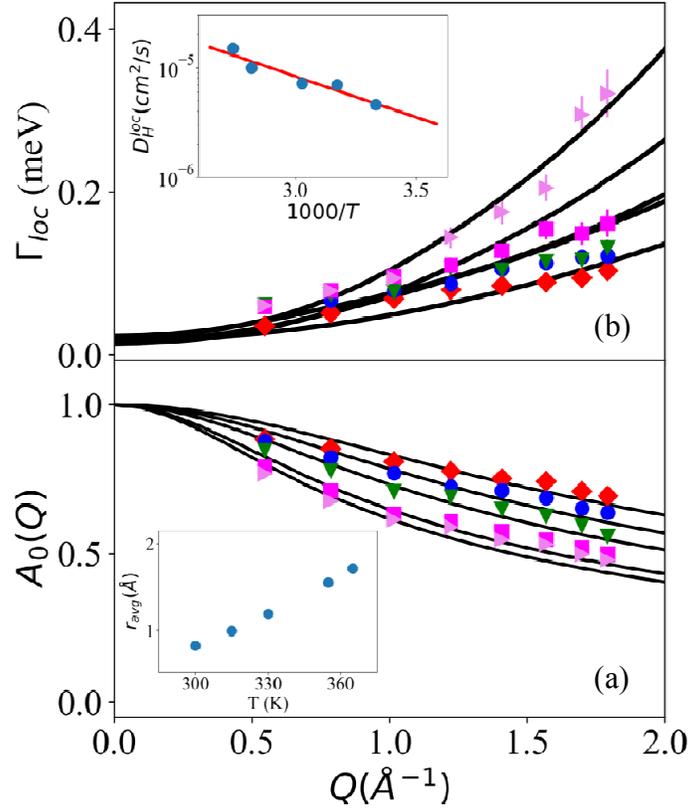

**Figure. 12** (a) *The EISF corresponding to localised motion in ADES obtained from fitting of the QENS spectra at all observed temperatures: 365 K(▶), 355 K(■), 330 K(▼), 320 K (●), 300 K (♦). The solid lines indicate the respective fits using eq. (9c) based on the modified LTD model. The variation of the average radius of localization with temperature is shown in the inset. (b) HWHM associated to the Lorentzian describing the localised translational diffusion of hydrogen atoms in ADES at all observed temperatures. The model function corresponding to LTD model is indicated by their respective solid lines. Arrhenius plot of the $D_H^{loc}$ is shown in the inset.*

### *4.4 Complexation of acetamide in ADES*

It has been observed from MD simulations and validated by QENS experiments that the long range jump diffusion process is significantly slower in ADES compared to ACM. In order to understand the origin of this, the molecular architecture of both liquids is studied using MD simulation trajectories. An acetamide molecule has hydrogen bond donor (nitrogen) and acceptor (oxygen) sites. The intermolecular pair distribution functions (PDF) of the hydrogen (associated to the amide group) and the oxygen atoms of the acetamide molecules, $g_{OH}(r)$, are calculated from MD simulation trajectories of ACM and ADES and shown in Fig. 13(a). The first peak observed at ~1.9 Å, correspond to the H-bonded acetamide-oxygen and amide-hydrogen atoms. The smaller area under the peak in the case of ADES compared to ACM indicates a decrease in the number of inter-amide H-bonding in ADES compared to ACM. The PDF of lithium ions and acetamide-oxygen in ADES, $g_{OLi}(r)$, is also calculated and shown in Fig. 13(b). The sharp peak which is also observed at 1.9 Å is an indication of strong H-bond association of acetamide-oxygen with the lithium ions, due to which inter-amide H-bonding is weakened in the ADES. The inset of Fig. 13(b) shows the PDF of amide-hydrogen ($H_A$) with the perchlorate-oxygen ($O_{Cl}$), where the strong first peak (~1.9 Å) is due to the H-bonding of the amide-nitrogen to the perchlorate ions in the ADES. These observations indicate that anion-acetamide and cation-acetamide complexes are formed in ADES. Dissociation of the salt in the system is necessary for the formation of these complexes. Generally, ionic dissociation and association of salt in a mixture is a process in dynamic equilibrium. However, this dynamic equilibrium can be shifted in the presence of acetamide which has H-bond donor and acceptor sites that have great affinity to the dissociated ions. Therefore, it can lead to the formation of complexes between ions and acetamide through hydrogen bonds[53]. From infrared and Raman spectroscopic studies on acetamide and LiTFSI eutectic mixtures, it has been reported that the oxygen site in acetamide has a great tendency to form a complex with the Li$^+$ anions[54]. Further, due to the association with lithium ions, the amide-nitrogen acquires a positive charge and tends to associate itself with the anion in the system. The ion-acetamide complexes formed in this process will have larger effective size and mass compared to acetamide molecules and hence lead to a slowing down of the diffusion process. A typical complex of acetamide with lithium and perchlorate in the system, observed in the MD simulation, is shown Fig. 13 (c). Therefore, the primary cause for the decrease of acetamide mobility in ADES can be associated to the formation of these complexes. Further, as the formation of these complexes also causes disruption of inter-amide H-bonds, it disallows the crystallization of both acetamide and salt and thereby causes a large depression in the freezing point of the ADES. Hence, the ion-acetamide complexes in the ADES can be considered as the key reason for, depression in its freezing point as well as slowing down of the diffusion of acetamide molecules. Since the long range diffusion process plays an important role in viscosity and conductivity of the liquid, the large viscosity and poor conductivity of the ADES can be ascribed to complexation of acetamide which significantly impedes the jump diffusion.

**Table 2**. *Dynamical parameters obtained from fitting of QENS spectra for the DES system at all the measured temperatures.*

| T (K) | $D_j$ ($10^{-6}$ cm$^2$/s) | $\tau$ (ps) | $r_{avg}$ (Å) | $D_H^{loc}$ ($10^{-5}$ cm$^2$/s) |
|---|---|---|---|---|
| 365 | 2.9 (±0.3) | 3.3 (±0.2) | 1.7 (±0.3) | 1.36 (±0.05) |
| 355 | 2.5 (±0.1) | 7.7 (±0.6) | 1.6 (±0.1) | 0.99 (±0.03) |
| 330 | 1.6 (±0.2) | 21.1 (±1.4) | 1.2 (±0.2) | 0.72 (±0.12) |
| 315 | 1.1 (±0.1) | 33.3 (±1.8) | 1.0 (±0.1) | 0.70 (±0.10) |
| 300 | 0.6 (±0.1) | 65.2 (±4.2) | 0.8 (±0.1) | 0.46 (±0.04) |

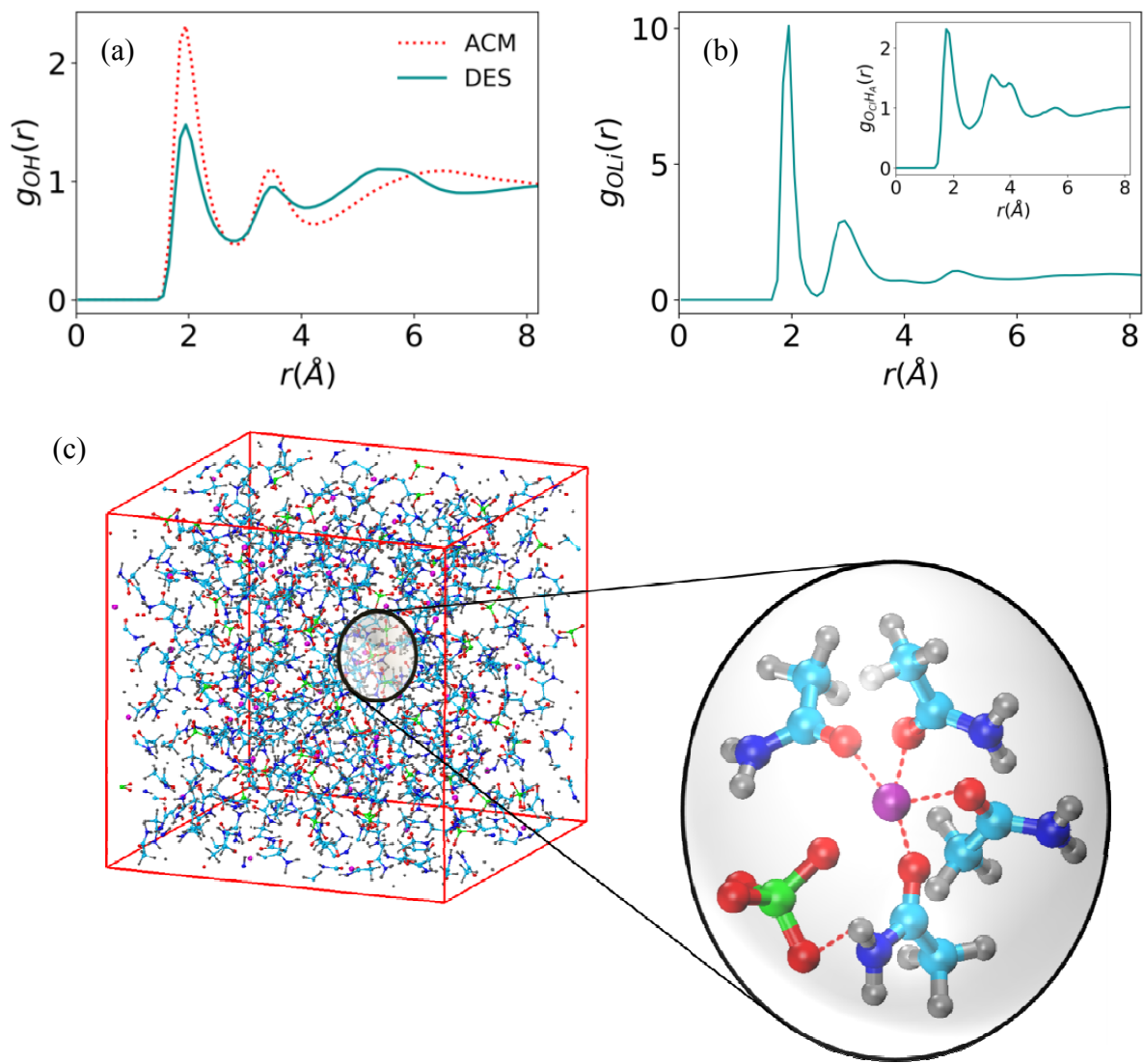

**Figure 13.** (a) *PDF of acetamide-oxygen with acetamide-hydrogen (of the amide group) in ACM and ADES. (b) PDF of acetamide-oxygen with lithium ions in ADES. (inset) PDF of amide-hydrogen with perchlorate-oxygen ions in the ADES (c) Snapshot of the simulation box and a complex formed between ions and acetamide through H-bonded network (hydrogen – gray, carbon – cyan, nitrogen – blue, oxygen – red, lithium – magenta, chlorine – green).*

## 5. Conclusions

A comprehensive study of the diffusion mechanism of acetamide in its molten phase and in a acetamide based deep eutectic solvent (ADES) prepared with lithium perchlorate, using neutron scattering and molecular dynamics (MD) simulation, is presented in this work. This work outlines an explicit method to construct a dynamical model for experimental analysis of relaxation processes in molecular liquids. With the atomistic insight obtained from MD simulations, a model for the molecular diffusion of acetamide is proposed involving a combination of localised diffusion within a transient cage and cage-to-cage jump diffusion. The analysis of neutron scattering data supports the suggested diffusion model in both molten acetamide and ADES. The nature of the diffusion mechanism is related to the extensive H-bond network in both the systems. The localised diffusion process is associated with the dynamics of the molecules while it is trapped by an H-bond with neighbouring molecules; on the other hand the jump diffusion process is associated to the diffusion of the molecule between two transient cages. The diffusion mechanism is found to be robust over a wide range of temperatures (300 K to 365 K) for ADES.

It is observed from MD simulation that the H-bond interaction between ions and acetamide facilitates the formation of ion-acetamide complexes in ADES. The increased size and mass of the complex causes decrease in the mobility of acetamide in the ADES compared to molten acetamide. In particular, the jump diffusivity is lower by at least a factor of three and the mean residence time between jumps is doubled. The complexes also inhibit crystallization and leads to a significant depression in the freezing point of the ADES. Hence, the ADES remains in the liquid phase even upto 300 K and the mechanism of the acetamide diffusion remains essentially unaltered in them. Both the jump and localised diffusion constants follow Arrhenius behaviour. While the knowledge of the long range diffusion process is vital to understand the mass transport properties like viscosity, conductivity etc. in the system, localised diffusion plays an important role in applications that involve confinement of the DES in the nanometer length scales. The large viscosity and conductivity of ADES can be attributed to the significantly slower long range jump diffusion in ADES compared to molten acetamide.

The results of this work serves as an important step in elucidating the mechanism of molecular diffusion in variety of deep eutectic systems based on metallic salts and amides. Since the diffusion of acetamide is strongly linked to the H-bond interactions in the ADES, one can engineer DES systems tailored to different transport properties by tuning their interaction strength.

**Acknowledgement:** We are grateful to Dr. Ranjit Biswas, S.N. Bose National Centre for Basic Sciences, Kolkata, India for fruitful discussion on DES systems. We would also sincerely thank Dr. Suman Das, University of Toronto, Canada and Dr. Niharendu Choudhury, Bhabha Atomic Research Centre, Mumbai, India for valuable suggestions regarding MD simulation. The neutron scattering experiments on DES and acetamide were conducted using the IRIS Spectrometer at the ISIS Neutron and Muon Source, and the digital object identifier for the data is https://doi.org/10.5286/ISIS.E.86392001.

# 6. References

1. D. R. MacFarlane, N. Tachikawa, M. Forsyth, J. M. Pringle, P. C. Howlett, G. D. Elliott, J. H. Davis, M. Watanabe, P. Simon and C. A. Angell, *Energy & Environmental Science*, 2014, **7**, 232-250.

2. K. N. Marsh, J. A. Boxall and R. Lichtenthaler, *Fluid Phase Equilib.*, 2004, **219**, 93-98.

3. A. Guerfi, M. Dontigny, P. Charest, M. Petitclerc, M. Lagacé, A. Vijh and K. Zaghib, *J. Power Sources*, 2010, **195**, 845-852.

4. H. Tadesse and R. Luque, *Energy & Environmental Science*, 2011, **4**, 3913-3929.

5. C. Arbizzani, M. Biso, D. Cericola, M. Lazzari, F. Soavi and M. Mastragostino, *J. Power Sources*, 2008, **185**, 1575-1579.

6. C. Wang, X. Luo, X. Zhu, G. Cui, D.-e. Jiang, D. Deng, H. Li and S. Dai, *RSC Advances*, 2013, **3**, 15518-15527.

7. G. Cevasco and C. Chiappe, *Green Chem.*, 2014, **16**, 2375-2385.

8. T. P. Thuy Pham, C.-W. Cho and Y.-S. Yun, *Water Res.*, 2010, **44**, 352-372.

9. M. Petkovic, K. R. Seddon, L. P. N. Rebelo and C. Silva Pereira, *Chem. Soc. Rev.*, 2011, **40**, 1383-1403.

10. A. P. Abbott, G. Capper and S. Gray, *ChemPhysChem*, 2006, **7**, 803-806.

11. A. P. Abbott, G. Capper, D. L. Davies, R. K. Rasheed and V. Tambyrajah, *Chem. Commun. (Cambridge, U. K.)*, 2003, DOI: 10.1039/b210714g, 70-71.

12. M. S. Sitze, E. R. Schreiter, E. V. Patterson and R. G. Freeman, *Inorg. Chem.*, 2001, **40**, 2298-2304.

13. Y.-F. Lin and I. W. Sun, *Electrochim. Acta*, 1999, **44**, 2771-2777.

14. A. P. Abbott, G. Capper, D. L. Davies, H. L. Munro, R. K. Rasheed and V. Tambyrajah, *Chem. Commun. (Cambridge, U. K.)*, 2001, DOI: 10.1039/b106357j, 2010-2011.

15. E. L. Smith, A. P. Abbott and K. S. Ryder, *Chem. Rev. (Washington, DC, U. S.)*, 2014, **114**, 11060-11082.

16. Q. Zhang, K. De Oliveira Vigier, S. Royer and F. Jérôme, *Chem. Soc. Rev.*, 2012, **41**, 7108-7146.

17. E. L. Smith*, *Transactions of the IMF*, 2013, **91**, 241-248.

18. A. P. Abbott, K. E. Ttaib, G. Frisch, K. S. Ryder and D. Weston, *Phys. Chem. Chem. Phys.*, 2012, **14**, 2443-2449.

19. A. Sun, H. Zhao and J. Zheng, *Talanta*, 2012, **88**, 259-264.

20. E. Gómez, P. Cojocaru, L. Magagnin and E. Valles, *J. Electroanal. Chem.*, 2011, **658**, 18-24.

21. M. Chirea, A. Freitas, B. S. Vasile, C. Ghitulica, C. M. Pereira and F. Silva, *Langmuir*, 2011, **27**, 3906-3913.

22. W. Liu, Y. Yu, L. Cao, G. Su, X. Liu, L. Zhang and Y. Wang, *J. Hazard. Mater.*, 2010, **181**, 1102-1108.

23. M. C. Gutiérrez, D. Carriazo, A. Tamayo, R. Jiménez, F. Picó, J. M. Rojo, M. L. Ferrer and F. del Monte, *Chemistry – A European Journal*, 2011, **17**, 10533-10537.

24. H. G. Morrison, C. C. Sun and S. Neervannan, *Int. J. Pharm.*, 2009, **378**, 136-139.

25. Y. Xie, H. Dong, S. Zhang, X. Lu and X. Ji, *Green Energy & Environment*, 2016, **1**, 195-200.

26. S. B. Phadtare and G. S. Shankarling, *Green Chem.*, 2010, **12**, 458-462.

27. Y. A. Sonawane, S. B. Phadtare, B. N. Borse, A. R. Jagtap and G. S. Shankarling, *Org. Lett.*, 2010, **12**, 1456-1459.

28. O. Coulembier, V. Lemaur, T. Josse, A. Minoia, J. Cornil and P. Dubois, *Chemical Science*, 2012, **3**, 723-726.

29. B. Guchhait, S. Das, S. Daschakraborty and R. Biswas, *J. Chem. Phys.*, 2014, **140**, 104514.

30. O. F. Stafford, *J. Am. Chem. Soc.*, 1933, **55**, 3987-3988.

31. A. Boisset, S. Menne, J. Jacquemin, A. Balducci and M. Anouti, *Phys. Chem. Chem. Phys.*, 2013, **15**, 20054-20063.

32. R. Chen, F. Wu, B. Xu, L. Li, X. Qiu and S. Chen, *J. Electrochem. Soc.*, 2007, **154**, A703-A708.

33. A. P. Abbott, *ChemPhysChem*, 2004, **5**, 1242-1246.

34. C. D'Agostino, R. C. Harris, A. P. Abbott, L. F. Gladden and M. D. Mantle, *Phys. Chem. Chem. Phys.*, 2011, **13**, 21383-21391.

35. M. Bée, *Quasielastic neutron scattering : principles and applications in solid state chemistry, biology and materials science*, A. Hilger, Bristol; Philadelphia, 1988.

36. H. Srinivasan, V. K. Sharma, S. Mitra and R. Mukhopadhyay, *J. Phys. Chem. C*, 2018, **122**, 20419-20430.

37. V. K. Sharma, S. Mitra and R. Mukhopadhyay, *Langmuir*, 2019, **35**, 14151-14172.

38. T. Burankova, R. Hempelmann, A. Wildes and J. P. Embs, *J. Phys. Chem. B*, 2014, **118**, 14452-14460.

39. J. P. Embs, T. Burankova, E. Reichert, V. Fossog and R. Hempelmann, *J. Phys. Soc. Jpn.*, 2013, **82**, SA003.

40. D. V. Wagle, G. A. Baker and E. Mamontov, *The Journal of Physical Chemistry Letters*, 2015, **6**, 2924-2928.

41. M. P. Allen and D. J. Tildesley, *Computer simulation of liquids*, Clarendon Press, 1989.

42. G. García, M. Atilhan and S. Aparicio, *J. Phys. Chem. C*, 2015, **119**, 21413-21425.

43. R. Stefanovic, M. Ludwig, G. B. Webber, R. Atkin and A. J. Page, *Phys. Chem. Chem. Phys.*, 2017, **19**, 3297-3306.

44. L. Martínez, R. Andrade, E. G. Birgin and J. M. Martínez, *J. Comput. Chem.*, 2009, **30**, 2157-2164.

45. R. B. Best, X. Zhu, J. Shim, P. E. M. Lopes, J. Mittal, M. Feig and A. D. MacKerell, *J. Chem. Theory Comput.*, 2012, **8**, 3257-3273.

46. J. N. Canongia Lopes and A. A. H. Pádua, *J. Phys. Chem. B*, 2004, **108**, 16893-16898.

47. I. S. Joung and T. E. Cheatham, *Journal of Physical Chemistry. B*, 2008, **112**, 9020-9041.

48. W. Smith, C. W. Yong and P. M. Rodger, *Mol. Simul.*, 2002, **28**, 385-471.

49. J. Qvist, H. Schober and B. Halle, *The Journal of Chemical Physics*, 2011, **134**, 144508.

50. K. S. Singwi and A. Sjölander, *Physical Review*, 1960, **119**, 863-871.

51. F. Volino and A. J. Dianoux, *Mol. Phys.*, 1980, **41**, 271-279.

52. A. G. Novikov, M. N. Robnikova and O. V. Sobolev, *Physica B: Condensed Matter*, 2004, **350**, E363-E366.

53. G. Berchiesi, G. G. Lobbia, V. Bartocci and G. Vitali, *Thermochim. Acta*, 1983, **70**, 317-324.

54. Y. Hu, Z. Wang, H. Li, X. Huang and L. Chen, *Spectrochimica Acta Part A: Molecular and Biomolecular Spectroscopy*, 2005, **61**, 2009-2015.